\documentstyle[preprint,seceq,epsf,epsfig,graphicx]{jpsj}
\def\runtitle{Energy Structure and
Adiabatic Magnetization Process of V$_{15}$}
\def\runauthor{Indranil {\sc Rudra}$^{a,b}$, K. {\sc Saito}$^b$, S. {\sc Ramasesha}$^a$ and
Seiji {\sc Miyashita}}
\newcommand{\mbold}[1]{\mbox{\boldmath $ #1 $}}
\def\bc{\begin{center}}
\def\ec{\end{center}}
\def\bfg{\begin{figure}}
\def\efg{\end{figure}}
\def\beq{\begin{equation}}
\def\eeq{\end{equation}}

\title{%
\textit
{\large  Characterization of the Energy Structure and 
Adiabatic Magnetization Process of  V$_{15}$}
}

\author{%
Indranil {\sc Rudra}$^{a,b}$, K. {\sc Saito}$^b$, S. {\sc Ramasesha}$^a$ and
Seiji {\sc Miyashita}\footnote{E-mail address:
miya@spin.t.u-tokyo.ac.jp}
}

\inst {
$^a$
Solid State \& Structural Chemistry Unit, Indian Institute of science, 
Bangalore 560 012, India\\
$^b$
Department of Applied Physics Graduate School of Science,
University of Tokyo, Bunkyo-ku Tokyo 113-8656
}

\recdate { \today }

\abst {
The energy structure of the V$_{15}$ is studied using exact
diagonalization methods, and the adiabatic changes of the
magnetization in the system with the sweeping field are investigated. 
We confirm that the Dzyaloshinskii-Moriya interaction leads to
an energy gap which allows the adiabatic change of the magnetization which has 
been found experimentally by Chiorescu et al., and also predict 
novel dynamics of the magnetization due to
this energy level structure.
}

\kword
{
V$_{15}$,adiabatic transition,Nanoscale magnets,Quantum dynamics
}

\begin{document}
\sloppy
\maketitle

\section{Introduction}

Interesting adiabatic change of magnetization in V$_{15}$ was 
discovered by Chiorescu, et al.,\cite{CWMBB} because of which the
energy level structure of V$_{15}$ has attracted much attention.
Detailed information of the energy levels is necessary in this case, since
time dependence of magnetization is almost completely governed by pure
quantum dynamics. Besides, we can also expect other interesting phenomena
related to quantum dynamics in this material, due to interesting energy level
structure that could emerge from new models.

The time-reversal symmetry and the half-odd value of the magnetization 
imply that the eigenstate at zero field must be at least doubly 
degenerate and therefore the quantum tunneling may be suppressed. 
However, this idea contradicts the observed adiabatic change 
of the magnetization in V$_{15}$ in so far as we consider only a single 
mode of magnetization. If the geometric frustration of interactions in 
V$_{15}$ is taken into account, one obtains a doubly degenerate avoided 
level crossing structure.\cite{Chiorescu-thesis,MN} 
That is, two avoided level crossing structures exist and 
they cross each other at $H=0$. The adiabatic change of the magnetization 
occurs in each avoided level crossing structure.\cite{Chiorescu-thesis} 
This degenerate structure has 
brought forward an interesting question: how does the magnetization change 
from 1/2 to 3/2? It has been shown, using a three spin model\cite{MN} that 
one of the states with $M=1/2$ evolves to a state of $M=3/2$ while the 
other state remains in the same $M=1/2$ state. 

\begin{figure}
$$
\begin{array}{c}
\epsfxsize=5.7cm \epsfysize=5.5cm \epsfbox{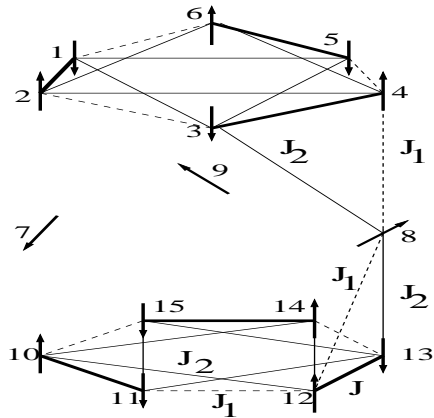}
\end{array}
$$
\caption{Structure of V$_{15}$: the bold line denotes $J$, the thin line $J_2$,
and
the dashed line $J_1$}
\label{V15-structure}
\end{figure}

Recently more realistic calculations have been done on the model which
includes all the 15 spins of V$_{15}$(Fig.\ref{V15-structure}).\cite{RRS,KC}
Experimentally there is evidence of a small splitting between the lowest
doublets of V$_{15}$. It is believed that Dzyaloshinskii-Moriya (DM) 
interaction is the cause of this gap.\cite{Chiorescu-thesis,MN,KC}
The V$_{15}$ molecule has a C$_3$ 
rotational symmetry and lacks a symmetry plane passing vertically through 
the middle of the molecule. That is, V$_{15}$ molecule does not have the 
symmetry between the right-hand rotation and left-hand rotation. 
This implies that the DM interaction in the $xy$ plane of the molecule 
can be non-vanishing. The DM interaction causes mixing of the two pairs of
low energy states of $S=1/2$ to form two sets of avoided level crossing 
states as a function of the external field $H$. Thus the system has an 
energy gap at $H=0$ and shows an adiabatic reversal of the magnetization.
Konstantinidis and Coffy\cite{KC} studied the energy structure of the 
system with the DM interaction by a perturbational method and obtained 
the dependence of magnetization on the strength of DM interaction. 

We extend their model by exploiting the C$_3$ symmetry and calculate the 
energy structure by exact diagonalization method.\cite{Davidson}
We employ the full 15-spin Hamiltonian with DM interactions, given by
\beq
{\cal H}={\cal H}_0+{\cal H}_{\rm DM}-H\sum_iS_i^z,
\eeq 
where
\beq
{\cal H}_0=\sum_{<ij>}J_{ij}\mbold{S}_i\cdot\mbold{S}_j
\eeq
and 
\beq
{\cal H}_{\rm DM}=
\sum_{<ij>}\mbold{D}_{ij}\cdot\left(\mbold{S}_i\times\mbold{S}_j\right) 
\eeq

\section{Method of Computation}

We have treated the DM interaction as a perturbation over the eight low-lying
states of the ${\cal H}_0$. The low-lying energy levels of ${\cal H}_0$ are 
very closely spaced with many degeneracies. To obtain the eigenvalues and 
eigenvectors with high accuracy and to gain information about the nature of 
the states we have used C$_3$ symmetry to divide the space into the doubly 
degenerate E and the nondegenerate A subspaces. We set up the Hamiltonian 
matrix of the Heisenberg Hamiltonian ${\cal H}_0$ in the symmetrized basis 
set in the chosen subspace. We obtain the eigenvalues and eigenvectors of the 
lowest eight energy levels using Davidson's algorithm.\cite{Davidson} 
In the next step, we
set-up the perturbation matrix ${\cal H}_{\rm DM}$ in the basis of the eight
low-lying eigenstates of  ${\cal H}_0$. This is achieved by setting up the
full matrix of  ${\cal H}_{\rm DM}$ in the full unsymmetrized basis and then
transforming the matrix into the basis of the eight low-lying eigenstates. 
To check the accuracy of the method, we have also diagonalized the total 
Hamiltonian calculated in the full unsymmetrized basis (where DM interaction 
is treated in a non-perturbative manner) with $D_x=D_y=D_z$=40 K and 
compared with the results from the above approximate method. We found not only 
the eigenvalues but also the eigenvectors of the lowest few states obtained by 
both methods to agree with each other very well for the lowest few eigenstates 
that could be obtained from the full Hamiltonian matrix. Therefore, we use the 
former method to calculate the quantities discussed in the following.

Using various sets of the parameters $(J,J_1,J_2,\mbold{D}_{ij})$, we have 
fitted the experimentally known energy level structure. In order to fit the 
energy gap between the ground state doublet and the first excited quartet 
in ${\cal H}_0$, we adopt the parameters\cite{RRS}
\begin{equation}
J=800{\rm K},J_1=54.4{\rm K},J_2=160{\rm K}.
\end{equation}
The configuration of the DM vectors $\mbold{D}_{ij}$ is not known yet.
At this moment we make use of the fact that the zero field energy gap of the 
avoided level crossing is about 80mK\cite{Chiorescu}, and the width of the 
magnetic jump from $M_z=1/2$ to $M_z=3/2$ is $\approx$ 0.75K\cite{CWMBBJMM}.
Although we have only two values to be fitted, there are many bonds in the 
molecule and therefore many DM parameters. To reduce the
set of DM parameters, we follow Konstantinidis and Coffy and assume that the 
$\mbold{D}_{ij}$ vector is nonzero only for the pair of spins between which 
the Heisenberg exchange $J$ is strongest. We further assume that the DM 
vectors of the upper hexagon and lower hexagons to be the same to minimize
the number of free parameters in the model.

\begin{figure}[hp]
\vspace*{-1cm}
\centerline{\includegraphics[height=10cm,width=16cm]
{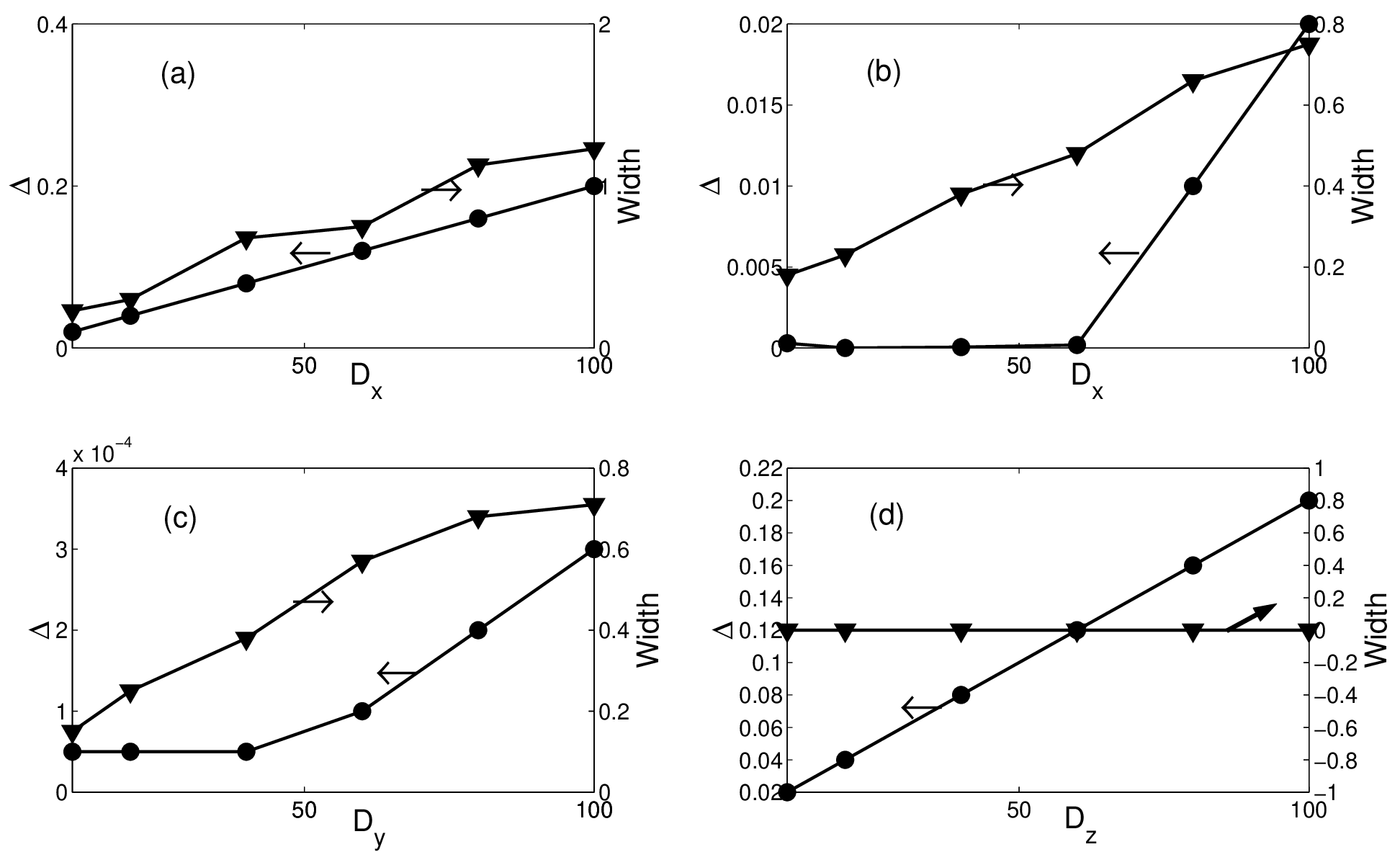}}
\centerline{\includegraphics[height=12cm,width=16cm]
{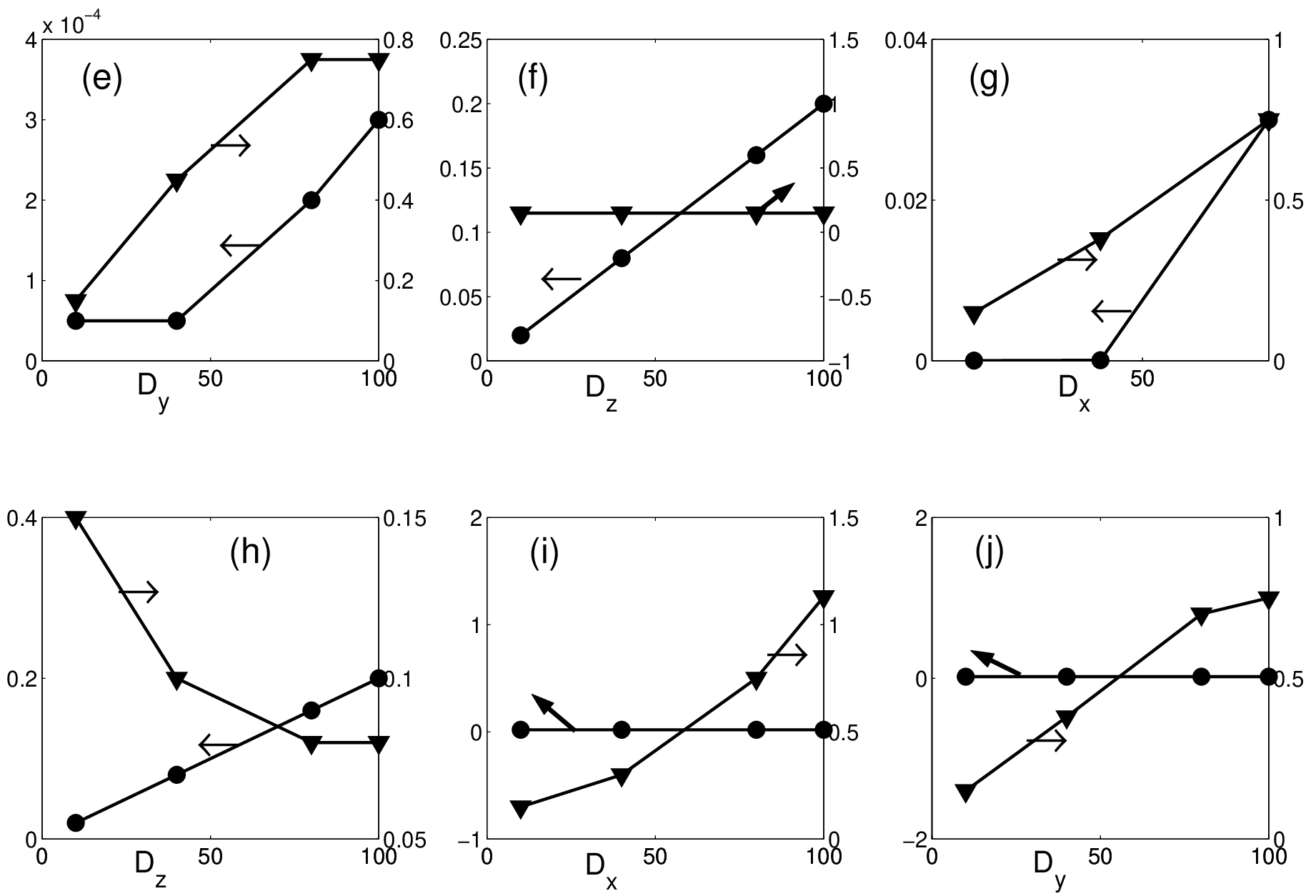}}
\end{figure}
\pagebreak
\vspace*{-1cm}
\begin{figure}[hp]
\centerline{\includegraphics[height=10cm,width=14cm]
{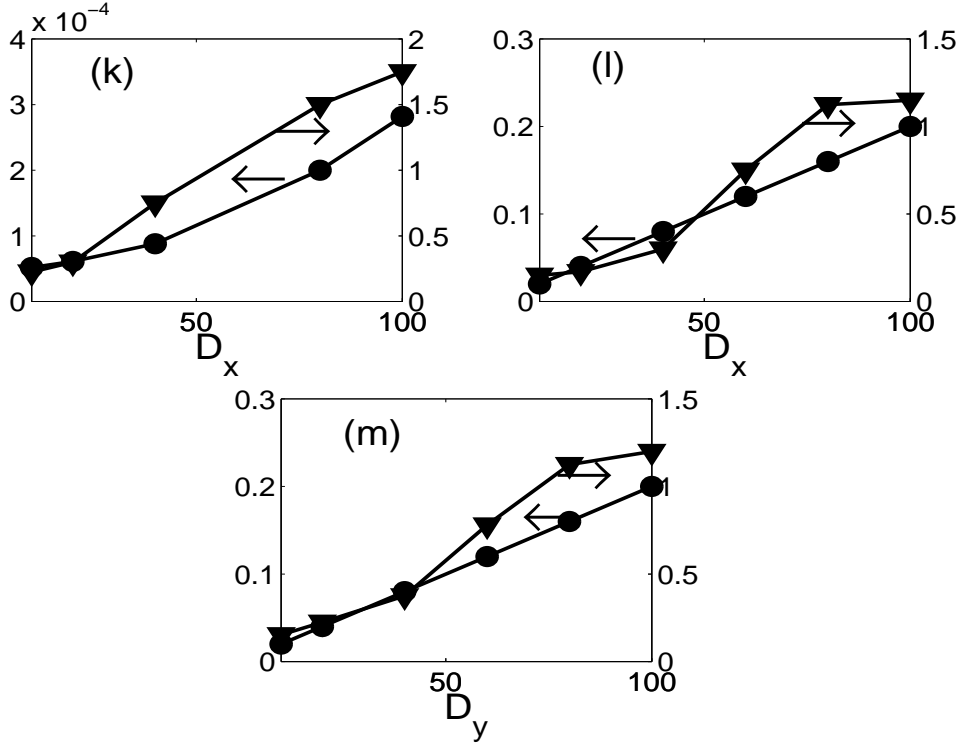}}
\caption{In all the figures Gap $\Delta$ (unit of Kelvin) is plotted on left $Y$
axis and Width (unit of Tesla) on right $Y$ axis. 
In Fig.(a) $D_x=D_y=D_z$.
In  Figs. (b), (c) and (d), we set $D_y=D_z=0, D_x= D_z=0$ and  $D_x= D_y=0$, 
respectively. 
$D_x$=10 K in Figs. (e) and (f). $D_y$=10 K in Figs. (g) and (h). 
$D_z$= 10 K in Figs. (i) and (j). 
In Figs. (e) to (j),
the component of DM vector, other than the
one taken as $x$-axis and the one fixed at 10 K, is set to 0 K. 
$D_x=D_y;D_z=0$ in Fig. (k). $D_x=D_z;D_y=0$ in Fig. (l). $D_y=D_z;D_x=0$ in Fig. (m).} 
\label{dval}
\end{figure}

We have explored the parameter space of $\mbold{D}_{ij}$ quite extensively
and show in Fig. \ref{dval} the variation of the energy gap between the
two doublets and the width of the transition at the level crossing
between $M$=1/2 and $M$=3/2. 
There, $(D_x,D_y,D_z)$ means the DM vector on 
the bond of $J$, say $\mbold{D}_{12}$. The DM vectors at other two bonds 
of a hexagon are given by a rotation of $120{^\circ}$:
\beq
\begin{array}{ccccc}
D_{34}^x={1\over2}\left(-D_x-\sqrt{3}D_y\right),& \quad&
D_{34}^y={1\over2}\left(\sqrt{3}D_x-D_y\right),& \quad&
D_{34}^z=D_z \\
D_{56}^x={1\over2}\left(-D_x+\sqrt{3}D_y\right),& \quad&
D_{56}^y={1\over2}\left(-\sqrt{3}D_x-D_y\right),& \quad&
D_{56}^z=D_z
\end{array}
\eeq

We note that the gap depends strongly on $D_z$
as we see in Fig.\ref{dval} (d), (f), (h), (l) and (m),
 and the width strongly on the components $D_x$ 
and $D_y$ as we see in Fig.\ref{dval} (b), (c), (e), (i), (j) and (k). 
Strong dependence of width on 
$D_x$ and $D_y$ can be understood from the fact that they contain terms which 
change the $M_z$ value of the state whereas $D_z$ part conserves the $M_z$ 
symmetry.
We found the set $(D_x,D_y,D_z)=(40,40,40)$
is a good candidate to reproduce the experimental gap and width.

We find the energy structure is similar to that of Model V in our previous 
study\cite{MN}. In Fig.\ref{E-V15}, the eight low-lying energy levels  are 
depicted as a function of the field.
\begin{fullfigure}
$$
\begin{array}{cc}[t]
\epsfxsize=6.7cm \epsfysize=5.5cm \epsfbox{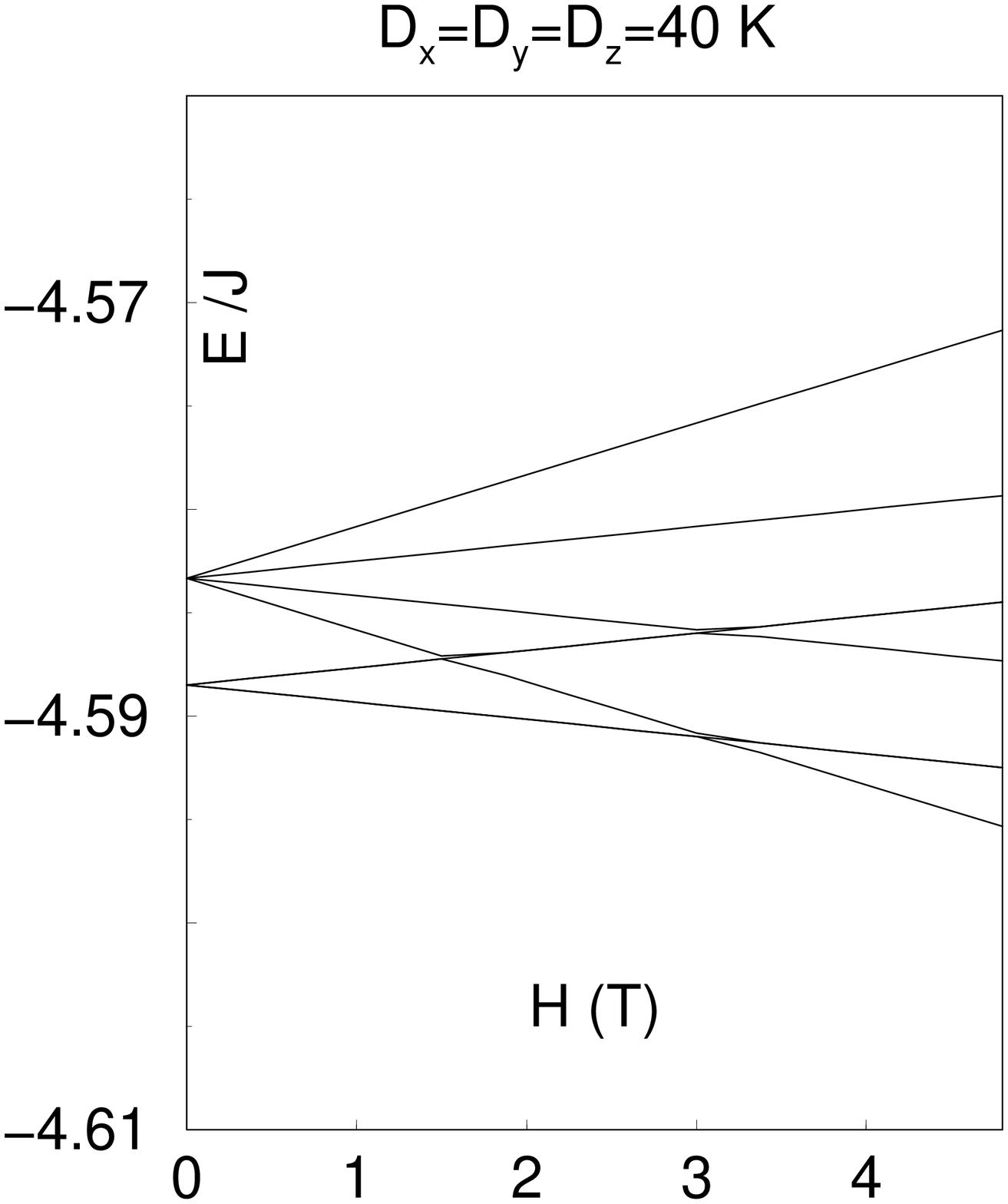}&
\epsfxsize=6.7cm \epsfysize=6.0cm \epsfbox{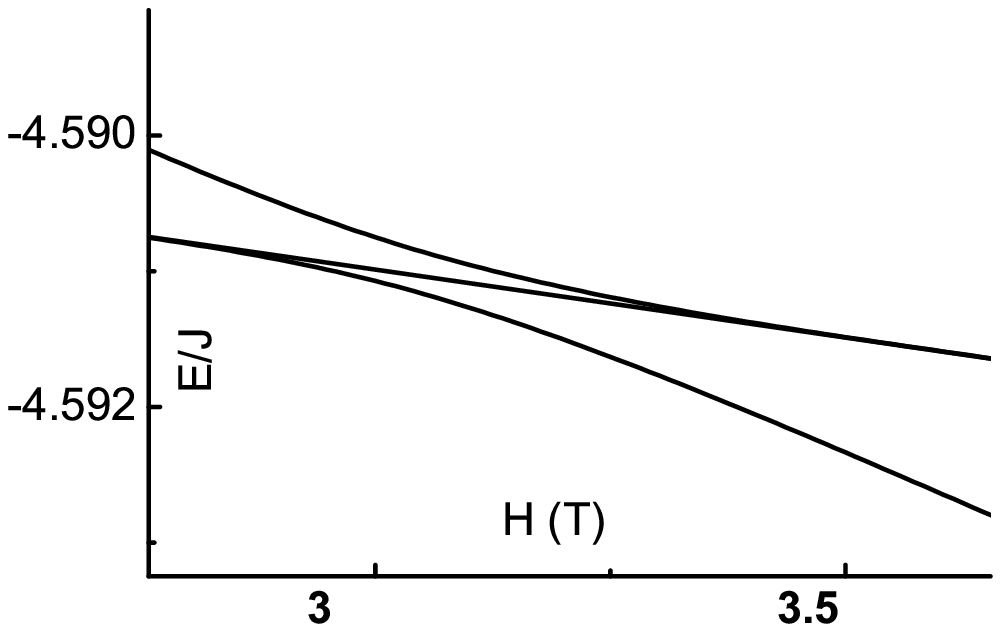}\\
{\rm (a)}& {\rm (b)}
\end{array}
$$
\caption{Energy levels obtained for V$_{15}$ (a) for the set of levels
S=1/2 and S=3/2. In (b) the avoided crossing between the M=1/2 and M=3/2 
levels is shown on a finer scale.}
\label{E-V15}
\end{fullfigure}
Around $H=0$ the low lying four states forms two sets of avoided level 
crossings. They are almost degenerate at small fields. However at the 
critical field $H_{\rm C}$ where the ground state magnetization changes from 
1/2 to 3/2, one of the degenerate states adiabatically changes to a state 
of $M_z\simeq 3/2$, while the other does not change the magnetization.
This structure is a consequence of the double-degeneracy of the $S=1/2$
ground states and causes interesting adiabatic process as discussed 
in the next section.

We find that a model with three spins interacting  antiferromagnetically
has the same symmetry of V$_{15}$,
and the model gives a good approximation for   
the present model if we choose $J=2.44$K and $D_y=0.06J$. 

\section{Adiabatic change of the magnetization}

The magnetization process of the present model shows 
a peculiar adiabatic magnetization process in a sweeping field.
Because the two avoided level crossing structures are very close to each other
near the zero field, we assume that the two low energy states are equally populated.
When the field is swept, one of them changes to a state of  $M=3/2$
and the other keeps $M=1/2$. Thus, in the adiabatic motion, the
magnetization changes as
\begin{equation}
{1\over2} \rightarrow \left({1\over2}+{3\over2}\right)\times{1\over2}=1
\end{equation}
instead of ${3\over2}$. 
We, therefore, expect that the magnetization changes to 1. This change 
will be observed at low temperature where the thermalization process is slow,
although the magnetization 1 finally relaxes to ${3\over2}$ due to the thermal effect. 
If we sweep back the field, the magnetization process will show a 
loop structure due to the thermalization effect. 
The process in a dissipative environments will be studied elsewhere \cite{disspation}. 
For very slow sweeping, the thermalization causes the system to 
realize the equilibrium magnetization process and  the magnetization 
directly changes to $M=3/2$.

\section{Hidden conservation law}

Here we would like point out another characteristic feature of the energy level 
structure. As we have pointed out in a simple model\cite{MN}, there 
exists a kind of selection rule. If we adiabatically change a state 
starting from the ground state at large negative 
field where the magnetization is almost $-3/2$, the state stays in the 
ground state until $H=0$. 
On the way the magnetization adiabatically changes from 
$-3/2$ to $-1/2$. At $H=0$ the state crosses with another state. Near $H=0$,
both states almost are degenerate and it is difficult to see the 
difference. The magnetization changes from $-1/2$ to $1/2$ around $H=0$. 
When we increase the field further, in the adiabatic change of the state, 
the magnetization does not change at the critical field $H_{\rm C}$ 
where the magnetization of the ground state changes from $1/2$ to $3/2$. 
That is, the magnetization adiabatically 
reached is $1/2$ for $H=+\infty$, although the ground state 
magnetization is $3/2$ for $H=+\infty$.

Here we adopt the following method to study the quantum dynamics of the 
model. We construct the total matrix ${\cal H}$ as mentioned previously
in the basis of lowest eight states but the field is now time dependent. 
We choose  the ground state at large negative field of $-6$ Tesla
as the initial state $|\psi_0>$. Then, the field is swept at a speed $v$ for 
a time $t$ up to a large positive field of 6 Tesla. We have studied the 
time evolution of the system by solving the time-dependent Schr\"odinger 
equation
\beq
i\hbar \frac{d\psi}{dt} = {\cal H}(t) \psi ~.
\eeq
The state is time-evolved by the equation
\beq
\psi(t+\Delta t) = e^{-i {\cal H}(t+\frac{\Delta t}{2}) \Delta t / \hbar} ~
\psi(t)~.
\eeq
The evolution is carried out by explicit diagonalization of the Hamiltonian
matrix ${\cal H}(t+\frac{\Delta t}{2})$, and using the resulting eigenvalues
and eigenvectors to evaluate the matrix of the time evolution operator 
$e^{-i{\cal H}(t+\frac{\Delta t}{2}) \Delta t/\hbar}$. Here we take 
$\Delta t=10^{-5}$, which is small enough because the maximum eigen value of the
energy is $E_{\rm max}=-4.598$ (in unit of J) and $E_{\rm max}\Delta t \ll 
O(1)$. 
We sweep the magnetic field with the velocity $v= 2.0\times 10^{-6}$, 
which is slow enough for the adiabatic change, because
the energy gap at $H=0$ is 80 mK and thus the Landau-Zener-St\"uckelberg
\cite{LZS} transition probability is estimated
   \begin{equation}
   p=1-\exp (- {\pi(\Delta E)^2 \over 2v}) = 0.007.
   \end{equation}

In Fig.\ref{Ad-mag}, the adiabatic change of the magnetization as a
function of $H$ is depicted for the 15 spin model. This property is also
found in the corresponding triangle model.
\begin{fullfigure}[t]
$$
\begin{array}{c}
\epsfxsize=5.7cm \epsfysize=5.5cm \epsfbox{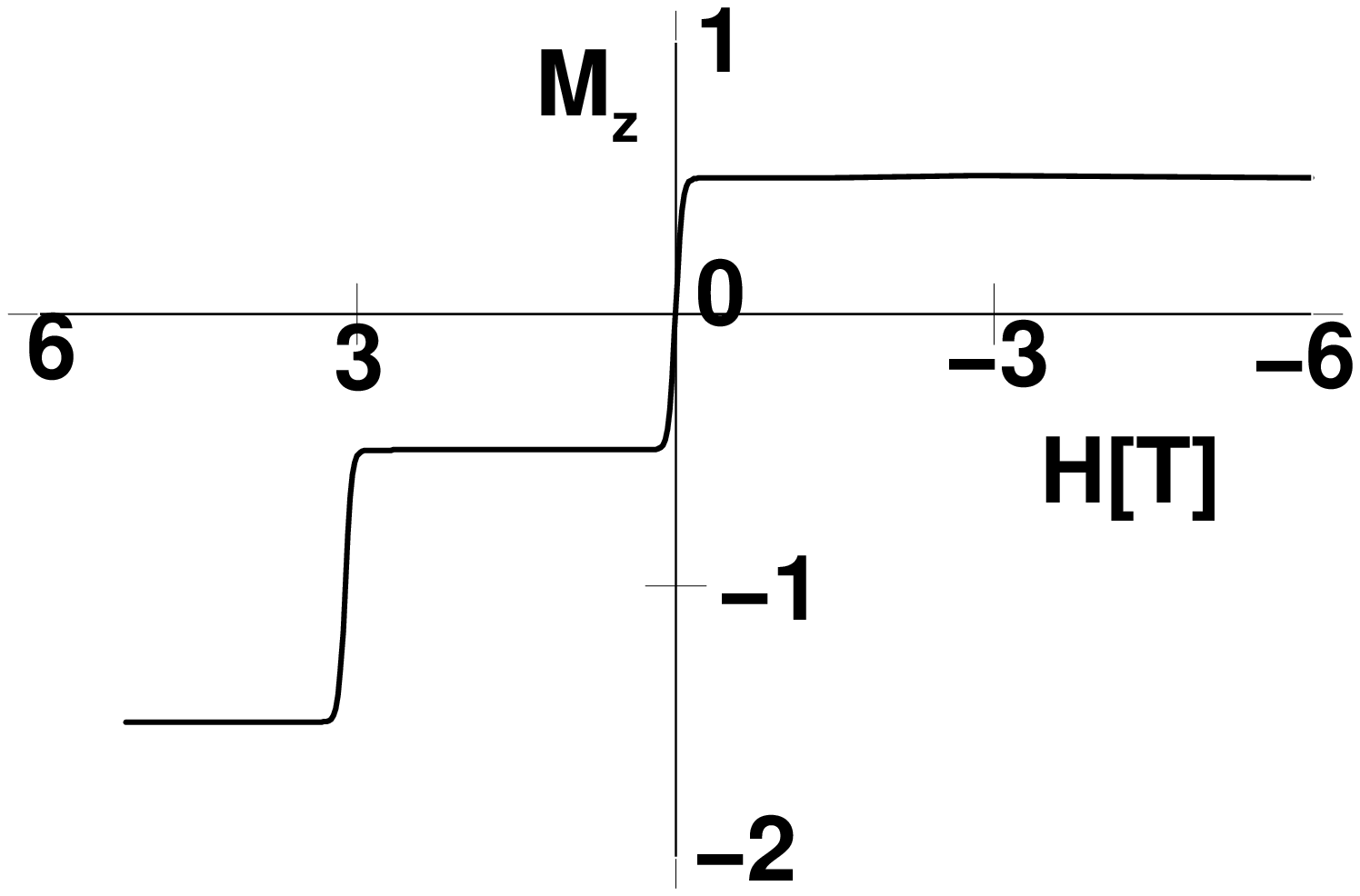}
\end{array}
$$
\caption{Adiabatic change of the magnetization starting
from the ground state with $M_z=-3/2$.}
\label{Ad-mag}
\end{fullfigure}

\section{Summary and Discussion}

We have investigated the energy level structure of V$_{15}$ using exact
diagonalization method. We find various peculiar properties 
of the energy level structure which are also expected from earlier studies of
an effective three spin model.\cite{MN}
First of all, we have shown that the DM exchange interaction leads to an energy
gap which allows for adiabatic change of the magnetization.\cite{CWMBB}
Furthermore, the two-fold degeneracy of the avoided level crossing structure
leads to an unusual energy level ordering in which the ground state 
magnetization changes from $1/2$ to $3/2$.

Because the adiabatic change is explicitly observed experimentally,
we expect that the novel dynamics of the magnetization due to 
this energy level structure can also be discovered in experiments.  
In this way we hope we can establish the role of real time quantum dynamics 
in magnetic systems, although we need a rather fast sweeping of the magnetic 
field in order to avoid the dissipative effects.\cite{Nojiri} 
Studies of  the dissipation effect will be reported elsewhere.\cite{disspation}.

\noindent{\bf Acknowledgements}

I.R. is grateful to Department of Science and Technology, India and Japan 
Society for the Promotion of Science for supporting his travel and stay in 
Japan. The present work is partially supported by a Grant-in-Aid from 
the Ministry of Education, Culture, Sports, Science and Technology
of Japan. We also acknowledge the super computer center of 
the solid state physics of University of Tokyo for the computational resource.

\end{document}